\pdfoutput=1
\documentclass[a4paper,american,aps,citeautoscript,floatfix,pdftex,pra,showpacs,superscriptaddress,twocolumn]{revtex4-1}
\usepackage{amsfonts,amsmath,amssymb}%
\usepackage{graphicx}%
\usepackage{hypernat}%
\usepackage[T1]{fontenc}%
\usepackage[utf8]{inputenc}%
\usepackage{textcomp}%
\usepackage{xspace}%
\usepackage{hyperref}%
\usepackage{appendix}

\newlength{\figwidth}
\newcommand{\degree}[1]{\xspace\ensuremath{^\circ}#1}%
\newcommand{\etal}{et al.}%
\newcommand{\ie}{i.\,e.}%
\newcommand{\ind}{indole}%
\newcommand{\indw}{indole(H$_2$O)}%
\newcommand{\Indw}{Indole(H$_2$O)}%
\newcommand{\indww}[1]{indole(H$_2$O)$_{#1}$}%
\newcommand{\Indww}[1]{Indole(H$_2$O)$_{#1}$}%
\newcommand{\invcm}{\ensuremath{\text{cm}^{-1}}}%
\newcommand{\ket}[1]{\ensuremath{\left|#1\right>}}%
\newcommand{\vt}{\ensuremath{v_\text{t}}}
\newcommand{\water}{H$_2$O}%

\newcommand{\cfel}{%
   \affiliation{Center for Free-Electron Laser Science, DESY, Notkestrasse 85, 22607 Hamburg, Germany}}
\newcommand{\uhh}{%
   \affiliation{University of Hamburg, Luruper Chaussee 149, 22761 Hamburg, Germany}}

\begin{document}
\title{Spatial separation of state- and size-selected neutral clusters}%
\author{Sebastian Trippel}%
\author{Yuan-Pin Chang}%
\cfel%
\author{Stephan Stern}%
\cfel\uhh%
\author{Terry Mullins}%
\author{Lotte Holmegaard}%
\cfel%
\author{Jochen K\"upper}%
\email{jochen.kuepper@cfel.de}%
\homepage{http://desy.cfel.de/cid/cmi}%
\cfel\uhh%

\date{\today}%
\begin{abstract}
   \begin{center}
      \textit{\normalsize Dedicated to Professor David W.\ Pratt on the occasion of his 75th birthday}
   \end{center}
   \noindent%
   We demonstrate the spatial separation of the prototypical \indw\ clusters from the various
   species present in the supersonic expansion of mixtures of indole and water. The major molecular
   constituents of the resulting molecular beam are \water, \ind, \indw, and \indww{2}. It is
   \emph{a priori} not clear whether such floppy systems are amenable to strong manipulation using
   electric fields. Here, we have exploited the cold supersonic molecular beam and the electrostatic
   deflector to separate \indw\ from the other molecular species as well as the helium seed gas. The
   experimental results are quantitatively explained by trajectory simulations, which also
   demonstrate that the quantum-state selectivity of the process leads to samples of \indw\ in
   low-lying rotational states. The prepared clean samples of \indw\ are ideally suited for
   investigations of the stereodynamics of this complex system, including time-resolved
   half-collision and diffraction experiments of fixed-in-space clusters. Our findings clearly
   demonstrate that the hydrogen bonded \indw\ complex behaves as a rigid molecule under our
   experimental conditions and that it can be strongly deflected.
\end{abstract}
\maketitle

\noindent%
The investigation of atomic and molecular aggregates -- so called clusters -- provides detailed
information for bridging the knowledge gap between microscopic atomic and molecular systems and
macroscopic samples. This includes, for example, the observation of ferroelectricity and
superconductivity~\cite{Moro:Science300:1265}, the investigation of proton
wires~\cite{Tanner:Science302:1736}, and the evidence for single water molecules acting as catalysts
of chemical reactions~\cite{Voehringer-Martinez:Science315:497}. Acid dissolution has been studied
at the microscopic level by stepwise attachment of solvents to molecules one by
one~\cite{Hurley:Science298:202}, and hydration shells are found to serve as heat sinks and to
enhance the photostability of DNA~\cite{Elsaesser:BioChem390:11}. Water clusters can have a
significant effect on the chemistry and the climate in earth's atmosphere~\cite{Vaida:JCP135:2}.

The actual dynamics of these processes could be investigated, for instance, using ultrashort-pulse
electron~\cite{Ihee:Science291:458, Siwick:Science302:1382} or X-ray
diffraction~\cite{Chapman:NatPhys2:839}, diffraction-from-within~\cite{Landers:PRL87:013002}, or
high-harmonic-generation experiments~\cite{Itatani:Nature432:867}. However, these experiments rely
on clean, homogeneous samples in order to produce so called ``molecular movies'' for individual
cluster sizes and isomers. Therefore, it is highly desirable to create pure samples of individual
cluster species.

For charged particles, size-selection can be performed according to the mass-to-charge ratio of the
clusters. Large ionic systems can also be separated according to their shape in ion-mobility
spectrometers~\cite{Helden:Science267:1483}. The full separation of structural isomers (conformers)
of specific sizes has so far been elusive, though. The preparation of state- and conformer-selected
samples of (small) neutral molecules has a long history going back to the seminal contributions by
Stern and Gerlach and by Rabi~\cite{Gerlach:ZP9:349, Rabi:PR55:526, Hamburg}. For small molecules,
static electric fields have extensively been used to select individual quantum
states~\cite{Reuss:StateSelection}. More recently, switched electric and magnetic fields have been
employed in the selection and slowing of individual or small sets of eigenstates of small
molecules~\cite{Meerakker:NatPhys4:595, Schnell:ACIE48:6010, Bell:MP107:99, Hogan:PCCP13:18705}. For
large molecules, dynamic (strong) focusing has to be employed~\cite{Auerbach:JCP45:2160,
   Bethlem:JPB39:R263}. The rotational state-selectivity of this technique has been
demonstrated~\cite{Wohlfart:PRA77:031404, Putzke:PCCP13:18962} and it has been used for the spatial
separation of individual conformers~\cite{Filsinger:PRL100:133003}. The
state~\cite{Holmegaard:PRL102:023001} and conformer~\cite{Filsinger:ACIE48:6900} selection of large
molecules has also been demonstrated using static inhomogeneous electric fields, going back to an
original proposal for state selection by Stern~\cite{Stern:ZP39:751}.

It is \emph{a priori} not clear, whether these techniques can be extended to complex molecules and
clusters with large numbers of internal degrees of freedom, many of which have low excitation
energies and will be excited even at low temperatures. Due to their large size (and correspondingly
large moments of inertia) and their floppiness these systems have very high densities of states and,
correspondingly, complex Stark energy curves with many real and avoided crossings. Indeed, it was
concluded before that the ``chaotic rotation'' of such complex asymmetric rotors makes their
manipulation very difficult~\cite{AbdElRahim:JPCA109:8507}. Here, we set out to demonstrate that the
electric deflection method is capable of spatially separating neutral polar molecules and clusters
according to their size and conformation, providing unprecedented possibilities to investigate the
structures and dynamics, even of complex aggregate systems.

We have chosen to demonstrate the spatial separation of the prototypical \indw\ cluster from the
various species in a supersonic cluster expansion of indole (C$_8$H$_7$N) and water (H$_2$O) seeded
in helium. Indole is the chromophore of the essential amino acid tryptophan and its intrinsic
properties have been widely studied. Its emmission properties are regularly used in fluorescence
studies of proteins, where spectral shifts are directly related to the chromophores'
environment~\cite{Vivian:BiophysJ80:2093}. The influence of water solvation on the indole
chromophore has been discussed extensively, since it has strong influence on the electronic states,
including the lowest-energy electronically excited states, which are considered to interchange their
energetic order with the addition of water and other polar molecules~\cite{Lami:JCP84:597}. The
structure of the \indw\ complex was determined in the ground and electronically excited
states~\cite{Korter:JPCA102:7211, Blanco:JCP119:880} and the influence of water on the lowest
electronic states discussed~\cite{Brand:PCCP12:4968:2010, Kuepper:PCCP12:4980, Kang:JCP122:174301}.
\Indww{2} clusters were also investigated using sophisticated multiple-resonance spectroscopic
techniques that showed, for instance, water-to-$\pi$-orbital binding
motifs~\cite{Carney:JCP108:3379}. The absorption spectra of indole in water solutions were
interpreted using strong shifts and reordering of the molecules' electronic
states~\cite{Lami:JCP84:597}.

Here, we experimentally demonstrate the preparation of a pure sample of the \indw\ clusters using
the electric deflector to separate this species from the cluster expansion \emph{soup}. We show
simulations that reproduce the obtained results and will allow predictions for the preparation of
further single-species samples even for such complex quantum systems.

\begin{figure}
   \centering%
   \includegraphics[width=\linewidth]{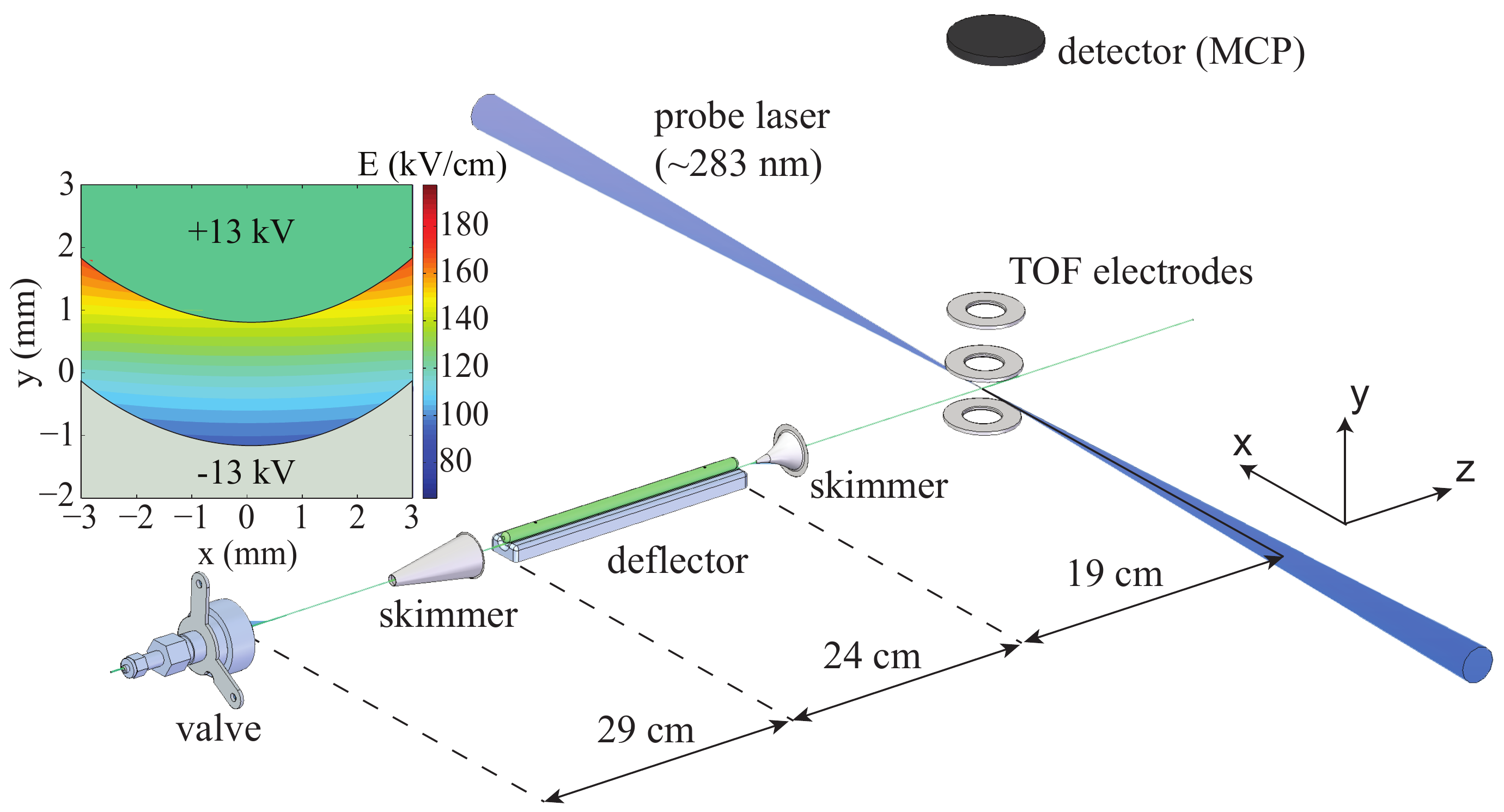}
   \caption{(Color online): The experimental setup consists of a pulsed molecular beam source, a
      deflector, a time-of-flight mass spectrometer and a pulsed dye laser. The inset shows a cut
      through the deflector and a contour-plot of the electric field strength for an applied voltage
      of 26~kV.}
   \label{fig:setup}
\end{figure}

The experimental setup is shown in \autoref{fig:setup}.
A supersonic expansion of a few mbar of indole and a trace of water molecules seeded in 50~bar of
helium is provided by a pulsed Even-Lavie valve~\cite{Even:JCP112:8068}. The valve is operated at a
temperature of 40\degree{C} and at a repetition rate of 20~Hz. Due to three-body collisions in the
early phase of the expansion \indww{n}\ clusters are formed. Under the employed conditions the
\indww{2} signal (density) is about an order of magnitude smaller than that of \indw. Larger
clusters are virtually absent from the beam. The resulting cluster beam passes a skimmer (Beam
Dynamics model 50.8) with an orifice diameter of 2~mm placed 22~cm downstream from the nozzle and
enters a differentially pumped vacuum chamber which houses the electric deflector. The length of the
deflection device is 24~cm and the vertical distance between the two electrodes, a stainless steel
trough and rod, is 2.3~mm. We typically apply voltage differences of 0~kV to 26~kV between the two
electrodes. This results in electric field gradients as shown in \autoref{fig:setup}, with field
strengths and gradients of up to 120~kV/cm and 250~kV/cm$^2$ in the center of the selector,
respectively. The electric field gradient exerts a force on the molecules and clusters by which they
are deflected towards the rod if they are in high-field-seeking states. Since the orifice diameter
of the first skimmer and the distance of the two electrodes are comparable, we have moved this
skimmer slightly towards the trough. This avoids the deflected beam crashing into the rod electrode
and improves the sensitivity to deflected molecules. After passing a second skimmer with an orifice
diameter of 2~mm mounted on a motorized translation stage the molecules enter the detection chamber.
\begin{figure}%
   \centering%
   \includegraphics[width=\linewidth]{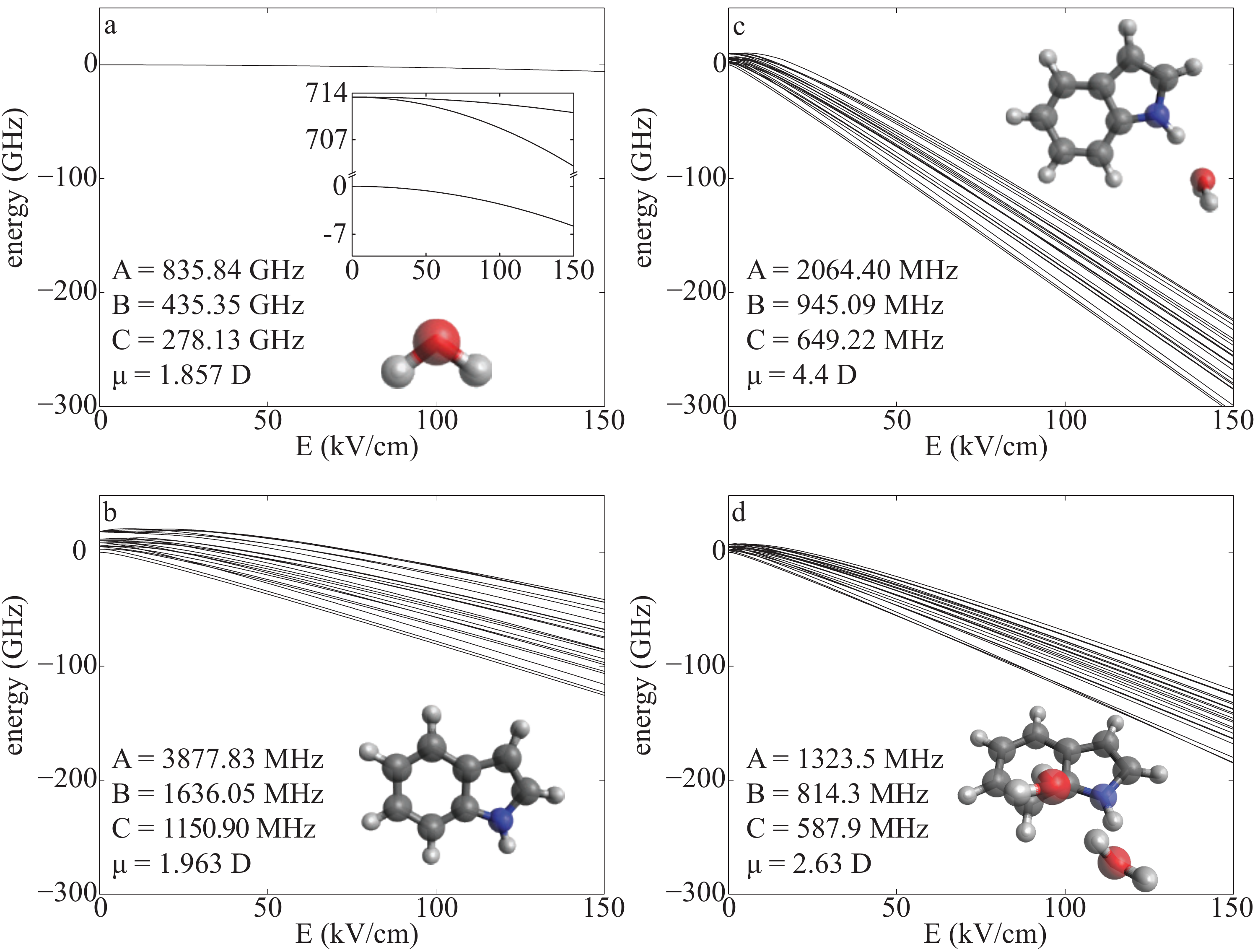}
   \caption{(Color online): Structures, molecular constants, and Stark energies of (a) \water, (b)
      \ind, (c) \indw, and (d) \indww{2}; see text for details.}
   \label{fig:molecules}
\end{figure}
This chamber houses a time-of-flight (TOF) mass spectrometer. The position of the second skimmer is
adjusted to maximize the transmission through the skimmer when the molecules are deflected. The
molecular beam is crossed by a laser at 90$^{\circ}$, in the center of the TOF-mass spectrometer.
Indole and indole-water clusters are ionized via one-color resonance-enhanced multi-photon
ionization (REMPI) at around 283~nm exploiting the lowest allowed electronic excitation of the
indole chromophore. Due to the resonant excitation step involving different laser frequencies, the
ionization is spectrally fully species selective. The laser light is provided by a nanosecond Nd:YAG
laser (Innolas) pumped frequency doubled dye laser (Fine Adjustment). The light is focussed by a
75~cm-focal-length lens and the typical pulse energy is a few mJ. For detection, the resulting ions
are accelerated towards a multi-channel plate (MCP) detector and time-of-flight (TOF) mass spectra
are recorded. Indole densities are determined from mass 117 (indole$^+$) and indole-water densities
are determined by summing masses 117 and 135 (indole$^+$ and \indw$^+$) following
(neutral-species-selective) resonance-enhanced ionization of the respective molecules or clusters.
The focusing lens is mounted on a translation stage and spatial density profiles of the molecular
beam are obtained by moving the lens perpendicular to the flight direction ($y$-coordinate, see
\autoref{fig:setup}) of the molecular beam.

\autoref{fig:molecules} shows structures, molecular constants, and Stark energies of all relevant
species. These are experimentally determined for \water~\cite{DeLucia:PRA5:487, Shostak:JCP94:5875},
\ind~\cite{Caminati:JMolStruct240:253, Kang:JCP122:174301}, and \indw~\cite{Blanco:JCP119:880,
   Kang:JCP122:174301} and based on \emph{ab initio} calculations
(GAMESS-US~\cite{Gordon:GAMESS:2005}, B3LYP/6-31+G*) for \indww{2}; the calculated structure and
dipole moment are the same as the ones reported by Carney \etal~\cite{Carney:JPCA103:9943,
   Somers:CP301:61}. The available experimental quartic centrifugal distortion constants for water,
\ind, and \indw\ are used in the calculations~\cite{Watson:VibSpecStruct6:1,
   Filsinger:JCP131:064309}. \label{centrifugal-distortion-not-necessary}However, the resulting
calculated deflection profiles using the rigid-rotor (not shown) and the semi-rigid-rotor
approximations are identical. The inset of \autoref{fig:molecules}\,a is a zoom into the Stark
energies of water in order to visualize its small Stark shifts. All populated states of the
molecules and clusters, at our experimental conditions, show a decrease of energy with increasing
field strength for the relevant range of $\sim$50--150~kV/cm. Thus, all molecules are in high-field
seeking states. The change in energy in the strong electric field depends on the specific rotational
quantum state. In general, the Stark effect decreases with increasing J quantum number giving rise
to the largest change in energy for the rotational ground states. Therefore, molecules in the lowest
rotational states are deflected the most. \Indw\ shows the largest Stark effect of all molecules and
clusters under consideration -- its potential energy changes 50~\% more than for \indww{2} and twice
as much as for \ind: If the clusters behave ``well'' (\emph{vide supra}), the biggest change in the
molecular beam profile, due to the interaction with the inhomogeneous electric field in the
detector, clearly is expected for \indw.

\begin{figure}
   \begin{center}
      \centering%
      \includegraphics[width=\linewidth]{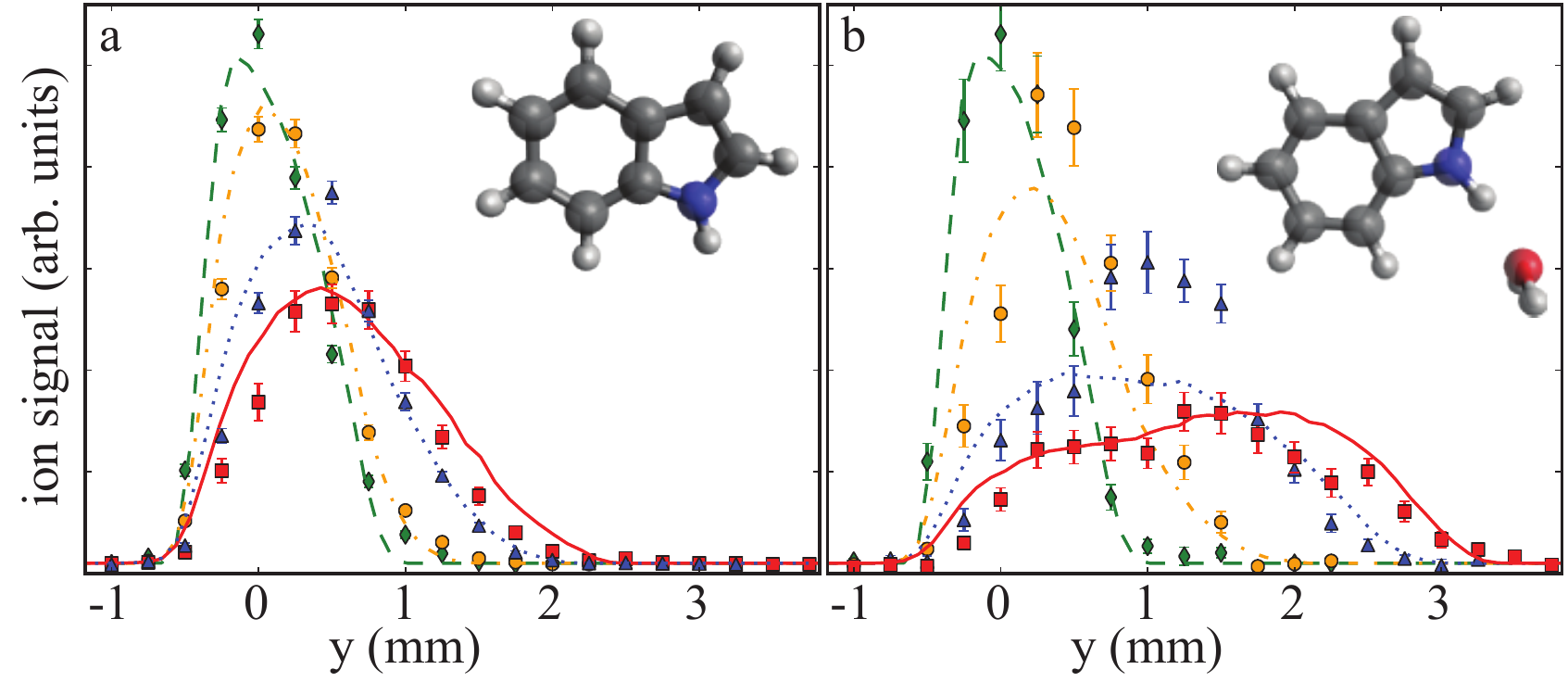}%
      \caption{(Color online): Deflection curves of (a) indole and (b) \indw\ for deflector voltages
         of $0$ (green diamonds), 10~kV (yellow circles), 20~kV (blue triangles) and 26~kV (red
         squares). The lines show the outcome of the respective trajectory calculations; see text
         for details.}
      \label{fig:deflection:separation}
   \end{center}
\end{figure}
\autoref{fig:deflection:separation} shows the \ind\ and \indw\ density profiles of the molecular
beam, obtained by scanning the focused REMPI laser beam perpendicular to the molecular beam
direction, for deflector voltages of 0, 10~kV, 20~kV and 26~kV. The width (FWHM) of the undeflected
beam is about 1~mm, limited by the mechanical aperture between the first skimmer and the deflector
trough in order to avoid collisions of the molecular beam with the rod and to increase the
sensitivity to detect deflected molecules. The density profiles show that all molecules and clusters
are deflected upwards and that they are dispersed as the deflector voltage is increased. This can be
attributed to all quantum states being high-field seeking for electric field strengths present in
the deflector and a variation (decrease) of the Stark effects with increasing rotational excitation,
respectively; see \autoref{fig:molecules}. The lines in \autoref{fig:deflection:separation}
correspond to simulated molecular beam profiles~\cite{Filsinger:JCP131:064309}. The force field
inside the deflector has been derived from the calculated electric field and the Stark curves and
the profiles have been obtained via Monte-Carlo trajectory simulations of individual molecules. The
good agreement between simulation and experimental data has been obtained by assuming a rotational
temperature of 4~K for \ind\ and of 6~K for \indw\ and by positioning the source and the two
skimmers accordant to the experimental conditions.

\begin{figure}
   \centering
   \includegraphics[width=\linewidth]{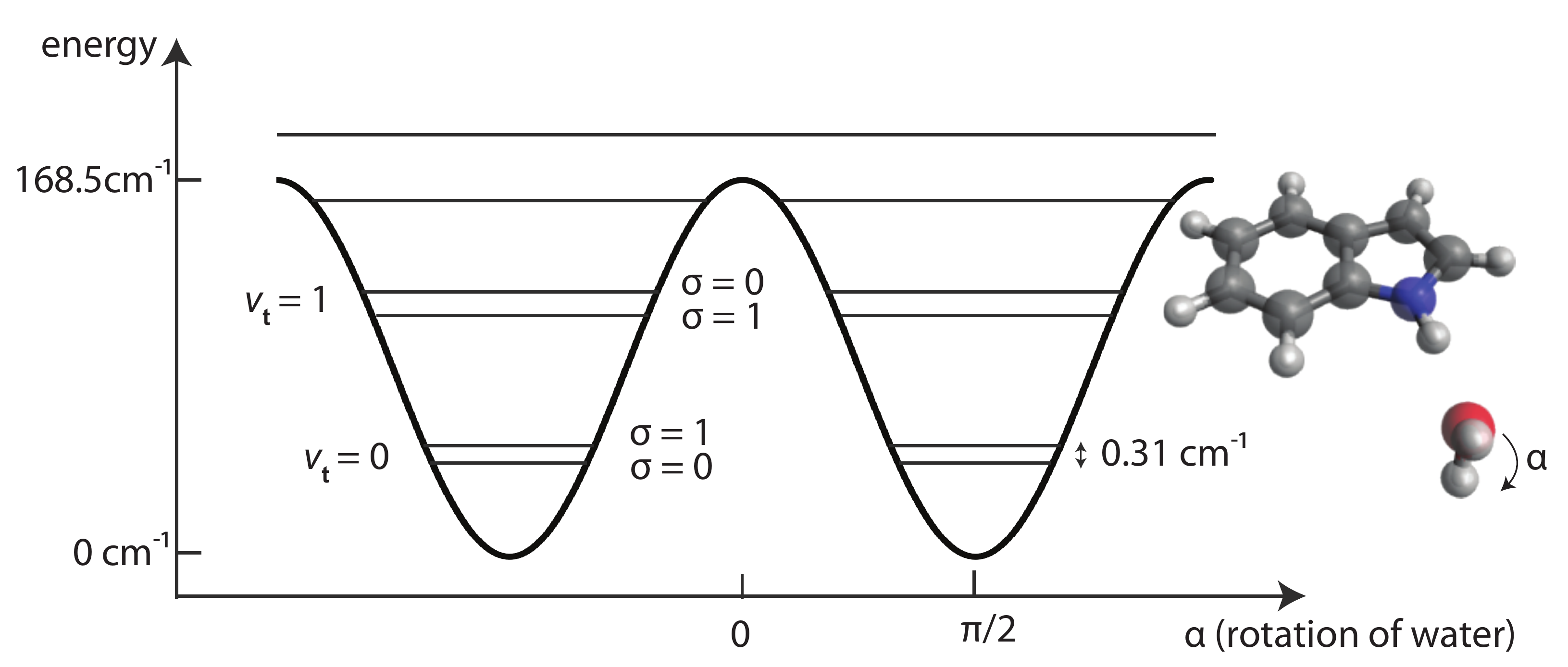}
   \caption{(Color online): The torsion potential and energy level scheme of \indw\ in its vibronic
      ground state (excluding torsion).}
   \label{fig:torsion}
\end{figure}

The \ind\ and water moieties of \indw\ are connected by a single hydrogen bond, and they can rotate
with respect to one another about this single bond. This torsional motion introduces the
complication that the cluster generally cannot be treated as a (semi)rigid rotor~\cite{semirigid}.
However, under our experimental conditions the cluster regains rigidity due to the low temperature
of our sample, the large rotational constant of the internal rotation of water, and the (dis)allowed
symmetry couplings of the populated states: Consider the twofold symmetric potential energy curve of
\indw\ shown in \autoref{fig:torsion}.
The angle between the molecular plane of \ind\ and the molecular plane of the water moiety is
denoted by $\alpha$. The torsional energy levels are labeled by the torsional quantum number \vt.
Each torsional level is split due to tunneling splitting and the corresponding torsional sublevels
are denoted by $\sigma=0,1$. The barrier height, $V_2$=168.5~\invcm, the angle of the axis of
rotation of the water moiety in the $bc$ symmetry plane of the cluster (corresponding to the $ab$
plane of bare \ind) with respect to the $b$ axis, $\beta$=55$^{\circ}$, as well as the rotational
constants have been taken from reference~\citealp{Korter:JPCA102:7211}. In order to determine the
energy levels we have numerically solved the one-dimensional Schrödinger
equation~\cite{Lewis:JMolStruct12:427, Gordy:MWMolSpec}. The included assumption that the rotating
moiety is a symmetric top molecule is not strictly fulfilled for water. However, the principal axes
of \indw\ do not change significantly when the water rotates and we can treat water as a symmetric
top~\cite{Gerhards:JCP104:967, Korter:JPCA102:7211}. With this model, we calculate a subtorsional
splitting in $\vt=0$ of $0.30~\invcm$ in agreement with the originally derived value of
0.31\,\invcm\ for this model~\cite{Korter:JPCA102:7211}. In order to estimate the contribution of
$\vt\ge1$ due to its initial state population prior to entering the strong field, we assume the
population of the energy levels to follow a Boltzmann distribution. The energy difference between
$\ket{\vt,\sigma}=\ket{0,0}$ and $\ket{\vt,\sigma}=\ket{1,1}$ is 70.9\,\invcm\ and the difference
between $\ket{\vt,\sigma}=\ket{0,1}$ and $\ket{\vt,\sigma}=\ket{1,0}$ is 76.5\,\invcm.
\label{vibrational-temperature}This gives rise to a relative population of $\vt=1$ with respect to
$\vt=0$ of less than 1\,\% at vibrational temperatures at 20~K or below, as is expected for our
supersonic expansion. Therefore, we neglect population with $\vt\ge1$ in our calculation, confirmed
by the good agreement with the experimental results.

The two subtorsional states in $\vt=0$ do not couple in the strong electric field of the deflector
due to their symmetry (see appendix~\ref{sec:symmetry-properties}). Coupling to $\vt\ge1$ can be
neglected, because the field in the deflector is not strong enough to efficiently couple $\vt=0$ to
these states (essentially because the energetic separation is too large). Therefore, under our
well-controlled experimental conditions \indw\ can be treated as a (quasi) rigid molecule.

\begin{figure}%
   \centering%
   \includegraphics[width=\linewidth]{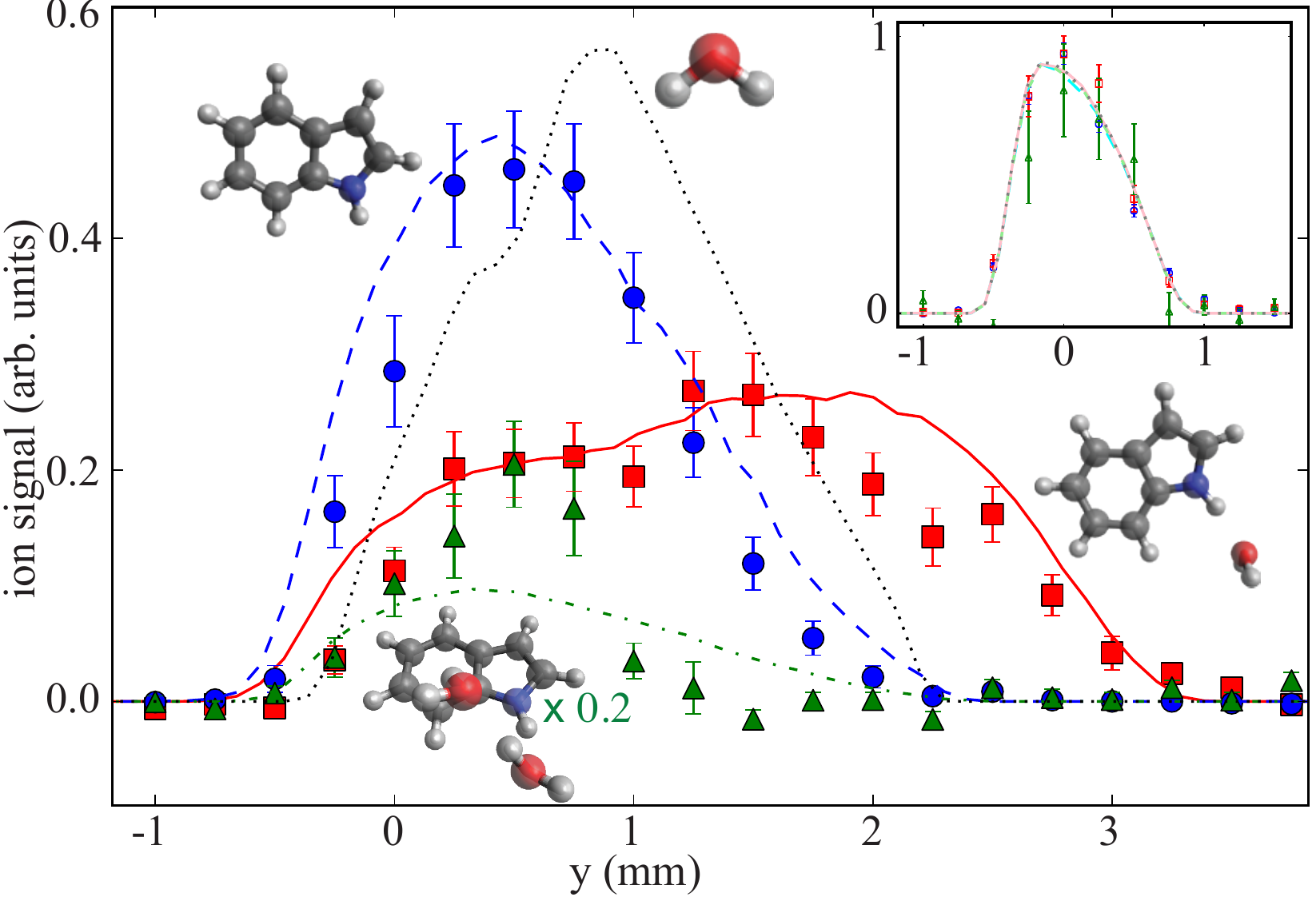}%
   \caption{(Color online): Normalized experimental (symbols) and simulated (lines) molecular beam
      profiles for water (dotted line), indole (blue circles/dashed line), \indw (red squares/solid
      line), and \indww{2} (green triangles/dash-dotted line) at a deflector voltage of 26~kV. The
      inset shows the corresponding undeflected normalized beam profiles.}
   \label{fig:deflection:voltage}
\end{figure}
\autoref{fig:deflection:voltage} shows the experimental and simulated beam deflection profiles for
indole, \indw, and \indww{2}, and the simulated profiles for water, for voltages of 0 and of 26~kV
applied across the deflector. The deflection-field-free density profiles are depicted in the inset
of \autoref{fig:deflection:voltage}. All undeflected and deflected beam profiles have been
normalized consistently to the field-free peak intensity for each species. This accounts for the
unknown (relative) cross sections for ionizing the molecules and clusters and for the unknown number
densities of the different species. The width of the undeflected beams is the same for all species.
For strong electric fields applied, the beam profiles of all species broaden -- due to the different
effective dipole moments of the populated states -- and shift upwards to larger $y$ values. \Indw\
is clearly more deflected than water, indole, and \indww{2}, as it is expected from the Stark curves
shown in \autoref{fig:molecules}. We note that ``exceptionally strong deflection'' was reported for
water in two of its $J=1$ states~\cite{Moro:PRA75:013415}. The reported behavior is reproduced by
our simulations, but this level of deflection is not very strong on the scale of our experiment.
These results, \ie, the profiles shown in \autoref{fig:deflection:voltage}, demonstrate that it is
possible to perform experiments on pure samples of \indw\ by simply selecting the species present in
the $y$-range from 2.3--3.3~mm. At $y=2.75$~mm (3.0~mm) the signal is 10~\% (4~\%) of the
undeflected peak intensity with only the 290 (110) lowest rotational quantum states populated,
instead of the 4600 states of \indw\ populated in the original molecular beam. Such state-selected
samples have been shown to be very amenable to alignment and orientation
control~\cite{Holmegaard:PRL102:023001, Nevo:PCCP11:9912}.

In conclusion, we have spatially separated \indw\ from a cluster beam containing water, indole, and
various indole-water clusters as well as atomic helium. We have experimentally demonstrated that
\indw\ is deflected more than the indole monomer and the \indww{2} cluster. We have quantitatively
simulated all experimental data using trajectory simulations. The good agreement of the simulations
with the experiments demonstrate that the manipulation of complex molecular systems and even weakly
bound clusters is possible and can be understood -- if the initial temperature of the sample is low
enough. These simulations also demonstrate that the \indw\ clusters are separated from \water\ and,
obviously, from the (undeflected) atomic helium seed gas as well as from larger water, indole, or
indole-water clusters, whose contribution to the beam is minimal and which have smaller
dipole-moment-to-mass ratios. Due to the state-specific deflection, it will also be possible to
separate different structural isomers, which will occur for larger
clusters~\cite{Carney:JPCA103:9943} -- analogous to the conformer separation of individual
molecules~\cite{Filsinger:PRL100:133003, Filsinger:ACIE48:6900}.

Recently, in a related experiment the selective alternating-gradient focusing of benzonitrile and
the benzonitrile-argon cluster was demonstrated~\cite{Putzke:JCP}.

The resulting pure samples of individual cluster species will allow for novel experiments
investigating the structures and dynamics of these species. One particular experiment we foresee is
the investigation of the energetics and stereodynamics, and their correlation, of half-collisions of
\indw\ in the molecular frame. One would 3D-orient~\cite{Nevo:PCCP11:9912} the molecules in the
deflected pure sample, photodissociate the cluster species using various photon energies, and,
successively, ionize the neutral water fragment from the reaction. The full momentum vector of the
water molecule in the molecular frame would then be determined using ion imaging. Similar
investigations using time-resolved gas-phase X-ray and electron diffraction experiments are also
envisioned~\cite{Filsinger:PCCP13:2076}. Such experiments would provide novel details about the
intermolecular interactions during, and their influence on, chemical reactions of complex
molecules~\cite{Suits:AccChemRes41:873, Townsend:Science306:1158, Mikosch:Science319:183}.

\begin{acknowledgments}
   We thank L.\ Gumprecht, J.\ Schulz, K.\ Długołęcki, H.\ Mahn, and H.\ Chapman for help regarding
   the setup of the experiment, the DESY infrastructure groups for support, and HASYLAB/Petra~III
   for hosting our laboratory. We acknowledge equipment loans from the Fritz Haber Institute,
   Berlin.
\end{acknowledgments}

\appendix
\section{Symmetry properties of \texorpdfstring{\indw}{indole-water} and coupling between
   sub-torsional levels}
\label{sec:symmetry-properties}
Here, we consider the symmetry properties of the torsional and sub-torsional levels of \indw\ and
the couplings between them due to an external dc electric field. The argumentation is equivalent to
the case of phenol(H$_2$O)~\cite{Berden:JCP104:972}. Because it is important regarding the
floppiness or rigidity of the \indw-cluster in our experiment the derivation is provided here in
detail.

\begin{table}[b]
   \begin{center}
      \hfill
      \begin{tabular}{c|cccc}
         \hline\hline
         G$_4$ & E & (12) & E$^*$ & (12)$^*$ \\
         \hline
         A$_1$ & 1 & 1 & 1 & 1 \\
         A$_2$ & 1 & 1 & -1 & -1 \\
         B$_1$ & 1 & -1 & -1 & 1 \\
         B$_2$ & 1 & -1 & 1 & -1 \\
         \hline\hline
      \end{tabular}
      \hfill
      \begin{tabular}{c|cc}
         \hline\hline
         G$_2$ & E & (12) \\
         \hline
         A & 1 & 1 \\
         B & 1 & -1 \\
         \hline\hline
      \end{tabular}
      \hfill\hspace{0pt}
      \caption{Character tables for the molecular symmetry groups G$_4$ and G$_2$~\cite{Berden:JCP104:972,
            Bunker:FundamentalsMolecularSymmetry}}
      \label{tab:character-tables}
   \end{center}
\end{table}
We describe the internal motion of the water moiety with respect to the \ind\ moiety using the
principal axis method (PAM)~\cite[page~582]{Lewis:JMolStruct12:427, Gordy:MWMolSpec}: We use the
principal axes of the cluster as a coordinate system in combination with an internally rotating
symmetric top moiety. The axis of internal rotation coincides with the symmetry axis of that
symmetric top and, therefore, internal rotation does not change the moments of inertia of the
overall rotation nor the cluster's principal axes. This condition is strictly speaking not fulfilled
for the case of \indw\ since the rotating water moiety is an asymmetric top molecule. However, the
moment of inertia of the rotating water molecule about its pseudo-symmetry axis is small compared to
the moment of inertia of the total molecule. The principal axes and moments of inertia of the
cluster are only slightly and insignificantly altered by the rotation of the water molecule. The
applicability of this approach is confirmed by related spectroscopic studies of hydrogen bound
molecular clusters~\cite{Gerhards:JCP104:967, Berden:JCP104:972, Korter:JPCA102:7211,
   Kang:JCP122:174301}.

In field-free space, \indw\ belongs to the molecular symmetry (MS) group $G_4 =\left\{E, (12), E^*,
   (12)^*\right\}$~\cite{Bunker:FundamentalsMolecularSymmetry} and the corresponding character table
is provided in \autoref{tab:character-tables}. The symmetry of the total wave function
$\Gamma_{tot}$ is of $B_1$ or $B_2$ symmetry since the total wave function must be antisymmetric
with respect to the exchange of the two protons (operation $(12)$). The symmetry of the electronic,
vibrational, torsional, rotational and nuclear spin states are summarized in
\autoref{tab:fourgroup:sym-table-indole_water}. In the last column, the resulting nuclear spin
statistical weights (nssw) are given.
\begin{table}
   \begin{center}
      \begin{tabular}{ccc|cccccccc}
         \hline\hline
         $\sigma$ & K$_a$ & K$_c$ & $\Gamma_e$ & $\Gamma_v$ & $\Gamma_t$ & $\Gamma_r$ & $\Gamma_t\otimes\Gamma_r$ & $\Gamma_{ns}$ & $\Gamma_{tot}$ & nssw\\
         \hline
         0 & $e$ & $e$ & A$_1$ & A$_1$ & A$_1$ & A$_1$ & A$_1$ & B$_2$  & B$_{2}$ & 1 \\
         0 & $o$ & $e$ & A$_1$ & A$_1$ & A$_1$ & A$_1$ & A$_1$ & B$_2$  & B$_{2}$ & 1 \\
         0 & $e$ & $o$ & A$_1$ & A$_1$ & A$_1$ & A$_2$ & A$_2$ & B$_2$  & B$_{1}$ & 1 \\
         0 & $o$ & $o$ & A$_1$ & A$_1$ & A$_1$ & A$_2$ & A$_2$ & B$_2$  & B$_{1}$ & 1 \\
         1 & $e$ & $e$ & A$_1$ & A$_1$ & B$_2$ & A$_1$ & B$_2$ & 3A$_1$ & B$_{2}$ & 3 \\
         1 & $o$ & $e$ & A$_1$ & A$_1$ & B$_2$ & A$_1$ & B$_2$ & 3A$_1$ & B$_{2}$ & 3 \\
         1 & $e$ & $o$ & A$_1$ & A$_1$ & B$_2$ & A$_2$ & B$_1$ & 3A$_1$ & B$_{1}$ & 3 \\
         1 & $o$ & $o$ & A$_1$ & A$_1$ & B$_2$ & A$_2$ & B$_1$ & 3A$_1$ & B$_{1}$ & 3 \\
         \hline\hline
      \end{tabular}
      \caption{The symmetry classification of eigenstates of field-free \indw\ and the respective
         nuclear spin statistical weights.}
      \label{tab:fourgroup:sym-table-indole_water}
   \end{center}
\end{table}

\begin{table}[b]
   \begin{center}
      \begin{tabular}{ccc|cccccccc}
         \hline\hline
         $\sigma$ & K$_a$ & K$_c$ & $\Gamma_e$ & $\Gamma_v$ & $\Gamma_t$ & $\Gamma_r$ & $\Gamma_t\otimes\Gamma_r$ & $\Gamma_{ns}$ & $\Gamma_{tot}$ & nssw \\
         \hline
         0 & $e$ & $e$ & A & A & A & A & A & B  & B & 1 \\
         0 & $o$ & $e$ & A & A & A & A & A & B  & B & 1 \\
         0 & $e$ & $o$ & A & A & A & A & A & B  & B & 1 \\
         0 & $o$ & $o$ & A & A & A & A & A & B  & B & 1 \\
         1 & $e$ & $e$ & A & A & B & A & B & 3A & B & 3 \\
         1 & $o$ & $e$ & A & A & B & A & B & 3A & B & 3 \\
         1 & $e$ & $o$ & A & A & B & A & B & 3A & B & 3 \\
         1 & $o$ & $o$ & A & A & B & A & B & 3A & B & 3 \\
         \hline\hline
      \end{tabular}
      \caption{The symmetry classification of eigenstates of \indw\ in a dc electric field and the
         respective nuclear spin statistical weights.}
      \label{tab:twogroup:sym-table-indole_water}%
   \end{center}
\end{table}

In an electric field the molecular symmetry group of \indw\ reduces to $G_2=\{E,(12)\}$, which
contains only two irreducible representations, $A$ and $B$; the character table is given in
\autoref{tab:character-tables}. The symmetries, under $G_2$, of all rotational quantum states of
\indw\ in its vibronic ground state are obtained directly by correlating $A_{1,2}$ and $B_{1,2}$
symmetries in $G_4$ to $A$ and $B$ in $G_2$, respectively. They are provided in
\autoref{tab:twogroup:sym-table-indole_water}. Torsional sublevels with the same quantum number
$\sigma$ have the same symmetry ($A$ for all $\sigma=0$ sublevels and $B$ for all $\sigma=1$
levels). The nonzero $\mu_a$ and $\mu_b$ components of the electric dipole moment are both of $A$
symmetry, because they do not change sign under the operation $(12)$, \ie, $(12)$ does not change
the dipole moment of the cluster. Thus, an electric field mixes only states having the same
$\sigma$. Consequently the sub torsional states within a given $\vt$ do not mix. In our experiment
more than 99\,\% of the population resides in the $\vt=0$ states, see
\autopageref{vibrational-temperature}, and \indw\ can be treated as a rigid rotor molecule under our
experimental conditions.

\bibliography{string,cmi,indole-water}%

\begin{thebibliography}{59}
\bibitem[{sem()}]{semirigid}
\bibinfo{note}{The ``semirigid'' refers to centrifugal distortion effects,
  which are known from high-resolution spectroscopic studies, but which are not
  necessary in our simulations,
  see~\autopageref{centrifugal-distortion-not-necessary}}
\end{thebibliography}
\bibliographystyle{jk-apsrev}%
\end{document}